\documentclass{iau}
\usepackage{graphicx}
\usepackage{xspace}
\usepackage{hyperref}

\title[360-degree videos of astrophysical simulations] 
{360-degree videos: a new visualization technique for astrophysical simulations}

\author[Christopher M.~P.~Russell]
{Christopher M.~P.~Russell$^1$}

\affiliation{$^1$X-ray Astrophysics Laboratory, NASA/Goddard Space Flight Center,\\ Greenbelt, MD 20771, USA (NASA Postdoctoral Program Fellow, administered by USRA)\\ email: {\tt crussell@udel.edu}
}

\pubyear{2017}
\volume{329}  
\jname{The lives and death-throes of massive stars}
\editors{J.J. Eldridge, L. Xiao, L.A.S. McClelland \\\& J.C. Bray, eds.}

\newcommand*{\SAs}{{Sgr\,A$^*$}\xspace}
\newcommand*{\thr}{{360$^\circ$}\xspace}

\begin{document}

\maketitle

\begin{abstract}
  360-degree videos are a new type of movie that renders over all 4$\pi$ steradian.  Video sharing sites such as YouTube now allow this unique content to be shared via virtual reality (VR) goggles, hand-held smartphones/tablets, and computers.  Creating \thr videos from astrophysical simulations is not only a new way to view these simulations as you are immersed in them, but is also a way to create engaging content for outreach to the public.  We present what we believe is the first \thr video of an astrophysical simulation: a hydrodynamics calculation of the central parsec of the Galactic centre. We also describe how to create such movies, and briefly comment on what new science can be extracted from astrophysical simulations using \thr videos.
\end{abstract}

\firstsection 
\section{Introduction}

\thr videos are a new type of video that displays over all 4$\pi$ steradian.  Spurred by the development of \thr cameras to capture \thr content, it is now possible to share these videos via sites like YouTube and Facebook, thus increasing their reach.

A natural application of this technology to astrophysics is to create \thr videos of simulations.  As opposed to traditional movies, which typically view the simulation domain from outside the simulation volume, or which move through the simulation domain but only render each frame in a predetermined direction, \thr videos immerse the viewers in the simulation and allow the viewers to choose where to look.  The three methods for viewing \thr videos are on a computer, with a hand-held smartphone or tablet, or in VR goggles.  For each method, you click and drag the video on the screen, pan the phone around, or simply look in different directions, respectively, to change the viewing orientation.  The VR goggles are by far the most immersive experience, but using a smartphone or computer makes these movies accessible to a much larger audience.

The \thr videos described here are not specifically VR content, in which the user puts on a VR headset plugged into a computer and can walk around to move through the simulation volume; the image in the goggles is then rendered in real time based on the user's position and viewing orientation.  On the other hand, sharable \thr videos have a predefined camera location chosen by the creator (just like with traditional movies), but the viewers choose where they look.

In this paper, we first describe how to create \thr videos from astrophysical simulations, then present what we believe
is the first such video shared online, which shows the distribution of material expelled by Wolf-Rayet stars in the Galactic centre, and finally discuss potential scientific applications of \thr videos.

\section{Constructing a 360-degree video}

The first step is to create a movie where the $x$- and $y$-axes are the azimuthal and polar axes, respectively.  Naturally, the $x$-axis should range from 0$^\circ$ to \thr, and the $y$-axis from 0$^\circ$ to 180$^\circ$, so the aspect ratio is 2-to-1 for the recommended square pixels.  Fig.\,\ref{fig1} shows a frame from the video presented here.

This is the most challenging step since most visualization software generates images where the $x$- and $y$-axes correspond to linear dimensions, not angular dimensions.  For our video, we modified the smoothed particle hydrodynamics (SPH) visualization program \texttt{Splash} (\cite[Price 2007]{Price07}) to render pixels across the full azimuthal and polar ranges at the desired angular width.  Once the images are created, the next step is to create a movie with a 2-to-1 aspect ratio using video creation software; we use \texttt{ffmpeg}.

The second step is to add metadata to the movie file that states that it is a \thr video file.  As we put our videos on YouTube, we followed the instructions here \url{https://support.google.com/youtube/answer/6178631}, which involves downloading a program called ``Spatial Media Metadata Injector.''  Importing the video, which must be in *.mov or *.mp4 format, to this program adds the necessary metadata.

The final step is to upload the video to YouTube.  Besides the aforementioned metadata, nothing special needs to be done to get the video into \thr format.  However, the usual YouTube tools to modify videos will overwrite this information, so they can not be used.  Therefore, the initial file into which the metadata is injected must be the final form.  Once uploaded, the \thr video is available for the world to see.

At present, the maximum resolution for a \thr video shared via YouTube is 2160s through the app and 4320s via a computer.  The `s' designation stands for `spherical,' and the number preceding it is the number of polar pixels; e.g., 2160s is 4320 azimuthal pixels by 2160 polar pixels.  The human eye with 20/20 vision can resolve $\sim$1$^\prime$, or $\sim$10800s, so the maximum resolution for \thr videos is $\sim$20-40\% of what humans can see.

\section{360-degree video of the Galactic centre}

\cite[Cuadra et al. (2008,]{Cuadra_etal08} \cite[2015)]{Cuadra_etal15} constructed numerical simulations of the 30 Wolf-Rayet (WR) stars and their winds orbiting \SAs within the central parsec of our galaxy.  Starting from the stellar locations 1100 yr ago,
the WRs orbit \SAs while ejecting their stellar wind material.  The central parsec quickly fills up with an ambient medium, into which the newly ejected wind material plows, causing wind-blown bubbles from the slow-moving stars and bow shocks around the fast-moving WRs.  The intent was to study the time-dependent accretion history of material onto \SAs (the WR winds are the dominant mass-injection source in the region), but was also successful in explaining the thermal X-ray emission resolved with \textit{Chandra} (\cite[Russell et al. 2017]{Russell_etal17}).

\begin{figure}[h]
\begin{center}
 \includegraphics[width=\textwidth]{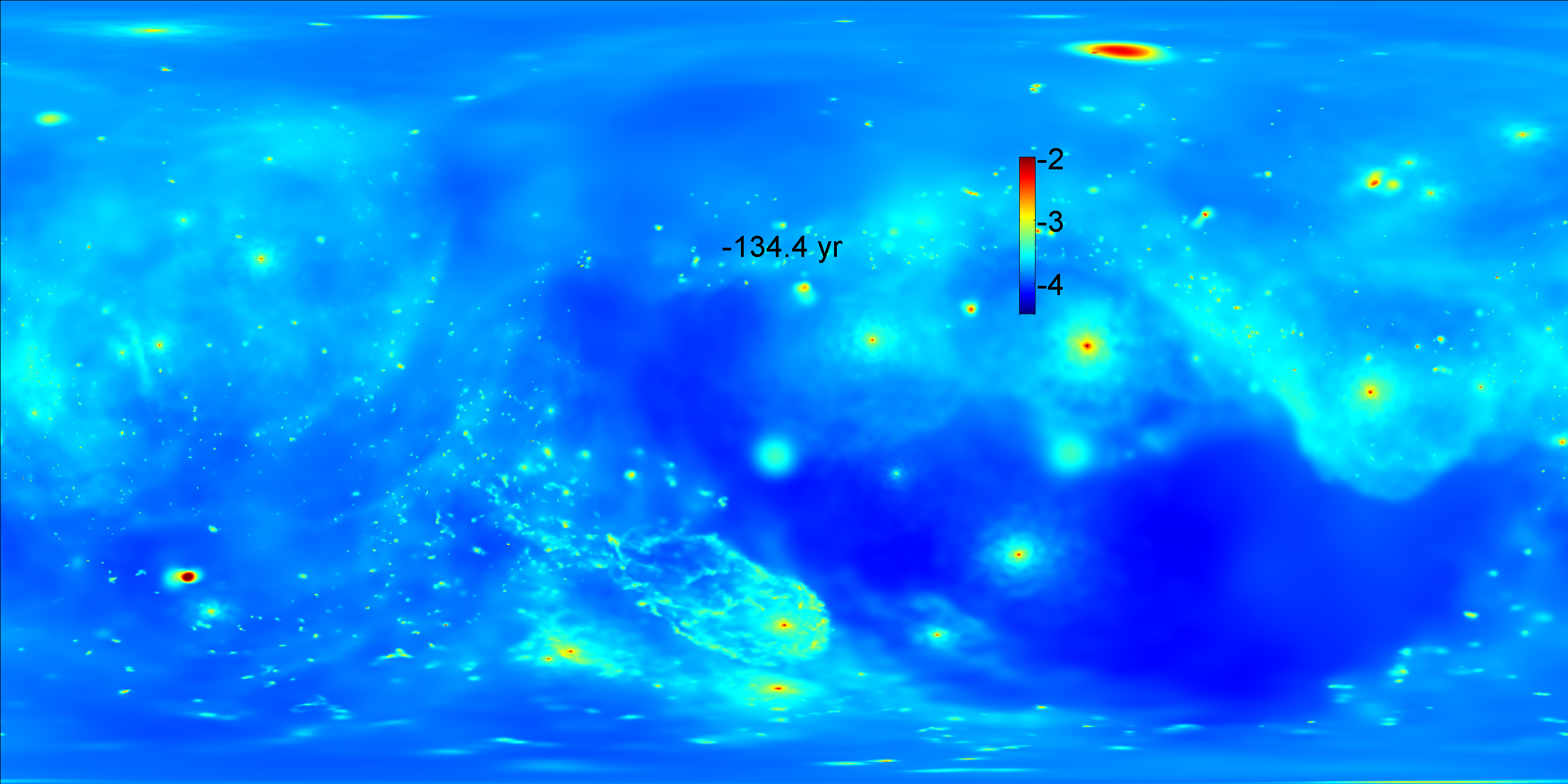}
 \caption{A single frame used to make the \thr video of the column density viewed from the exact centre of our galaxy.  The $x$-axis goes from 0-360$^\circ$ in azimuthal, while the $y$-axis is 0-180$^\circ$ in polar.  Adding the metadata tells YouTube to warp this into a sphere when viewing the movie.}
   \label{fig1}
\end{center}
\end{figure}

\begin{figure}[h]
\begin{center}
 \includegraphics[trim={0 7cm 0 7cm},clip,width=5.77cm]{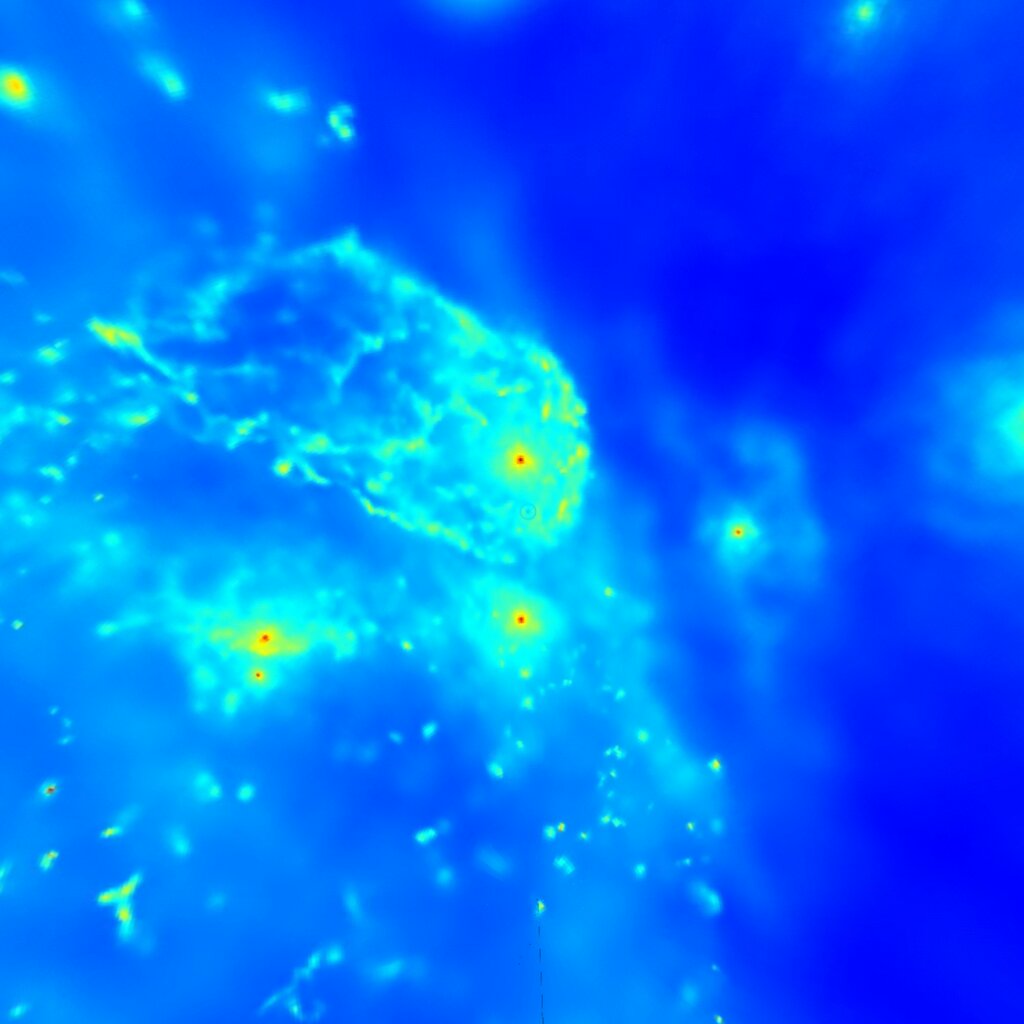}
 \caption{A screenshot of the movie when viewed from inside VR goggles.
 }
   \label{fig2}
\end{center}
\end{figure}

We created a \thr video from an updated hydrodynamic simulation of these 30 WR stars and their winds by rendering column density from the position of the centre of the simulation, i.e. at the location of \SAs.  The link to the video is \url{https://youtu.be/pK59iu4cNRM} \footnote{Alternatively, it might be better, particularly if viewing on a smartphone or VR goggles, to go to CMPR's YouTube channel -- either \url{http://tinyurl.com/cmpr360video}, or search for ``Christopher Russell astronomy'' in the app -- and select the video ``Galactic Center Column Density.''}.  Fig.\,\ref{fig2} shows the snapshot of Fig.\,\ref{fig1} viewed in \thr format.  Note that the distortion evident in Fig.\,\ref{fig1}, which is due to the image being polar vs.\ azimuthal angle and the viewing region being near the bottom of the plot, is gone when viewed in the \thr format of Fig.\,\ref{fig2}.

To our knowledge, a shorter version of this video, published to YouTube on 15 Nov.\ 2016, is the first \thr video of an astrophysical simulation ever created and shared.  (This video was published 10 days later.)

\section{Future work}

We are currently studying which science insights can be obtained from viewing simulations in this \thr manner. For example, we have produced a new movie with 10$\times$ higher time sampling (\url{https://youtu.be/BWiBIol7gzQ}).  This affords a clear view of the inspiraling and stretching of clumps as they plummet towards \SAs. Additionally, a \thr movie of the colliding wind binary $\eta$ Carinae shows when the primary wind completely engulfs the secondary star around periastron passage (\url{https://youtu.be/RzF6u0on_tw}). More applications will certainly be developed in the future.

\end{document}